\documentclass[epsfig,12pt]{article}
\usepackage{epsfig}
\usepackage{graphicx}
\usepackage{array}
\usepackage{color}
\usepackage{bm}
\usepackage{cancel}
\usepackage{bm}
\usepackage[nottoc,notlof,notlot]{tocbibind} 
\usepackage[titles]{tocloft}

\newcommand{\beq}{\begin{equation}}   
\newcommand{\eeq}{\end{equation}}
\newcommand{\beqn}{\begin{eqnarray}}   
\newcommand{\eeqn}{\end{eqnarray}}

\newcommand{\gsim}{\lower.7ex\hbox{$
\;\stackrel{\textstyle>}{\sim}\;$}}
\newcommand{\lsim}{\lower.7ex\hbox{$
\;\stackrel{\textstyle<}{\sim}\;$}}
\usepackage{showlabels}

\begin{document}
\begin{flushright}
FTPI-MINN-22-27,\quad UMN-TH-4134/22
\end{flushright}
\begin{center}

\vspace{3mm} 

{\Large The Inception, the Concept and the Second Life of Supersymmetry}

\vspace{0.4cm}

Mikhail Shifman

\vspace{0.2cm}

{\em William I. Fine Theoretical Physics Institute, University of Minnesota, Minneapolis, MN 55455}

shifman@umn.edu

\vspace{5mm}

{\large Abstract}
\end{center}

I give a general non-technical review of supersymmetry and its modern applications in phenomenology and quantum field theories at strong coupling.
Invited Talk at the conference  {\sl Frontiers of Fundamental Physics, FFP16, May 23-26, 2022, Istanbul, T\"urkiye}

\vspace{5mm}

When I received  invitation to give a talk at this conference I hesitated between two options:  discussing some recent works of mine in nonperturbative supersymmetry or presenting a broad review of the subject beginning from its inception. Given the present situation in high energy physics (HEP) -- that it currently undergoes a dramatic change -- and given that there are so many young people in the audience, I decided in favor of the broad review. At the end of the talk I will just mention in passing one new result not yet reported at other conferences.

\begin{figure}[t]
\centerline{\includegraphics[width=9.5cm]{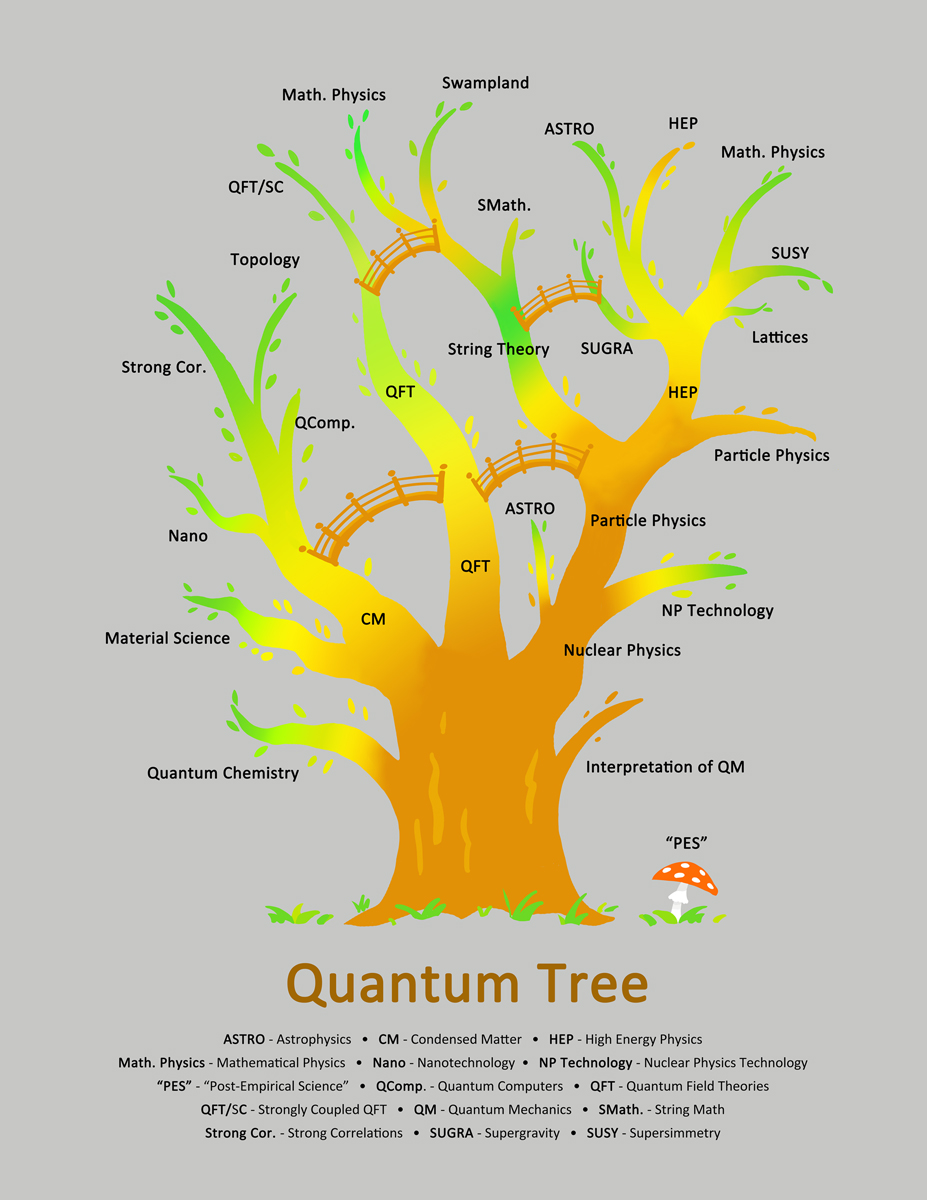}}
\caption{\small 
Quantum tree.
} 
\label{QT} 
\end{figure}

I will start with a  sketch of the {\em quantum tree} (Fig. \ref{QT}), which I designed to give a general idea of how quantum theory developed. The so-called ``old" quantum mechanics of Planck, Bohr, and Sommerfeld emerged in the beginning of the 20th century as an attempt to understand natural phenomena, e.g. black body radiation spectrum. Universal quantum mechanics of today was created by Heisenberg and Schr\"odinger in 1925; thus, the world will celebrate its centenary soon. This revolutional theoretical discovery gave rise to three main branches of the quantum tree. Quantum field theory -- the focus of my talk -- followed immediately. In 1926, Born, Heisenberg, and Jordan \cite{BHJ} turned their attention to  electromagnetic field in the empty space. 
They gave a formula for the electromagnetic field as a
Fourier transform and used the canonical commutation
relations to identify the coefficients in this Fourier
transform as operators that destroy and create photons. The inception of the first ``practical"  quantum field theory -- four-dimensional quantum electrodynamics -- can be traced back  to a 1927 paper of Paul Adrian Maurice Dirac \cite{PAMD}.  The first paper in \cite{PAMD} is entitled ``The Quantum Theory of the Emission and Absorbtion of Radiation." Namely here Dirac coined the name quantum electrodynamics (QED). The names of Born, Heisenberg, and Dirac are of course known to every student. This is not the case with Jordan, a rising star in theoretical physics in the late 1920s.\footnote{ Well... mathematicians are aware of Jordan algebras. } Why? The reason is that he became active in the Nazi movement very early. He traded his divine talent for a hateful ideology. For him physics faded away. This is a lesson for young people: never trade your talents for dubious (to put it mildly) causes. 
Let us return to the quantum tree. Some branches you see are green, some are light brown. The difference is self-evident -- the former continue to grow 
while the latter are more or less in the dormant state. 

The right side of the tree started from nuclear physics (NP) and quantum field theory (QFT).  In the 1950s, NP grew into particle physics which eventually evolved into high energy physics (HEP), Standard Model (SM) and supersymmetric (SUSY) phenomenology, lattices, astroparticle physics and other topics.
At its early stages, QFT was represented primarily by QED, a weakly coupled theory. Today it developed an impressive array of methods to deal with strong coupling in multiple applications. String theory also branched off from HEP. Since 1980s it had an enormous impact on the community but, unfortunately, failed to meet initial expectations
of ``theory of everything." A huge and most important in practical life branch is on the left. Quantum  condensed matter theory which started in the 1930s gave rise to all of quantum chemistry, material science, nanotechnologies, and theory of strongly correlated and topological systems. It presents a basis for current intense efforts aimed at quantum computing. Needless to say, all branches are densely intertwined and cross-fertilize each other. I managed to show only four bridges. In fact, their number is much larger.

On the right side at the bottom of the tree you see a  fly agaric mushroom, red with white stalks. It is poisonous. More exactly, in small doses it is hallucinogenic. The berserkers used it before battles to induce their trancelike state. It depicts a  tendency which goes under the name post-empiric science (PES) first discussed in earnest at a conference titled {\sl Why Trust a Theory? Reconsidering Scientific Methodology in Light of Modern Physics} in 2015 in Munich. To my mind, this is a fruitless and even dangerous development. Since Galileo's time primary in physics was  experimental study of natural phenomena. Theories were based on experiment. Further experiments were then used either to confirm a given theory or falsify it. Instead, PES suggests to use other non-empiric criteria to this end, such as self-consistency, endorsement by leading experts in the area, etc. 

The arrow in the middle indicates the birth of supersymmetry 50 years ago. The first publication of four-dimensional supersymmetric field theory belongs to Golfand and Likhtman \cite{GL}. Golfand was a member of the Theory Department at Lebedev Physics Institute in Moscow, Likhtman was his student. If you open the paper you will see that it was submitted to the Editors on March 10, 1971.\footnote{Due to specific Soviet conditions of total censorship this means that the work was completed a few months before. } They reported two major discoveries. First, they extended the Poincar\'e algebra to super-Poincar\'e, introducing conserved supercharges,
\beq
\left\{\bar{Q}_{\dot\alpha} Q_{\alpha} \right\} = 2P_{\dot\alpha  \alpha}=2\left(\sigma^\mu\right)_{\dot\alpha  \alpha}P_\mu\,.
\label{1}
\eeq
Here $Q\,\,, \bar Q$ are spinorial supercharges which anti-commute  producing  the energy-momentum vector  $P_\mu$ .
 The Coleman-Mandula theorem  \cite{CM}
stating  that in dynamically nontrivial theories,
i.e. those with a nontrivial $S$ matrix, no geometric extensions of
the Poincar\'{e} algebra are possible, was published in 1967. The spinorial extension of Poincar\'{e}
algebra (\ref{1}) was discovered as a ``loophole'' in the above
 theorem.

In the very same paper Golfand and Likhtman presented the first supersymmetric model in four dimensions, super-QED, which they constructed by a painful matching of a large set of coefficients.\footnote{This fact is less known in the literature. Golfand and Likhtman included in their  consideration a non-vanishing photon mass term which is allowed by renormalizability in Abelian gauge theories. Those who are interested in the early history of supersymmetry can turn to the collection \cite{KS}.}

I should mention a curious feature: there is only one reference in \cite{GL}, to Schweber's textbook \cite{sch}. The authors had nothing else to refer to. Everyone can be proud of such degree of novelty. 

By the way, Likhtman was the first to construct representations of the superalgebra (\ref{1}); he also noted that (\ref{1})  implies vanishing vacuum energy density. I refer the reader to his extremely interesting talk
\cite{Lik}.

Next, we will pass to Wess and Zumino whose seminal paper appeared in 1973 and caused a revolution in quantum field theory.
But first I will jump to 1974 and present the simplest supersymmetric gauge field theory generalizing super-QED -- non-Abelian Yang-Mills \cite{FZ,SS}. Its Lagrangian (usually referred to as SUSY-YM or supersymmetric gluodynamics\footnote{The latter term was coined by A. Polyakov.}) can be written as follows,
\beq
{\mathcal L}=-\frac{1}{4g^2}G_{\mu \nu}^{a}
 G_{\mu\nu}^{a}+\frac{i}{g^2}\lambda^{a\alpha}\mathcal{D}_{\alpha\dot{\beta}}\bar{\lambda}^{a\dot{\beta}}
\label{2}
\eeq
where $G_{\mu \nu}^{a}$ is the gluon field strength tensor and $\lambda$ is a Weyl fermion field -- gluiono -- in the adjoint representation of the gauge group.
Both gluon and gluino have two physical components per each value of the color index $a$. The conserved
spin-$\frac{3}{2}$ current $J_{\beta}^{\mu}$ connecting gluons to gluinos has the form (in
spinorial notation)
\begin{equation}
J_{\beta\alpha\dot{\alpha}}\equiv(\sigma_{\mu})_{\alpha\dot{\alpha}}J_{\beta}^{\mu}=\frac{2i}{g^{2}}G_{\alpha\beta}^{a}\bar{\lambda}_{\dot{\alpha}}^{a}\,.
\label{3}
\end{equation}
The supercharges are defined as
\beq
Q_\alpha = \int d^3\vec{x} \,J^0_\alpha (t,\vec{x}),\quad \bar{Q}_{\dot\alpha} =\left(Q_\alpha\right)^\dagger\,.
\eeq
Acting on bosons they produce fermions and vice versa.

Julius Wess and Bruno Zumino spent the year of 1973 at CERN. Their motivation and the path to supersymmetry were totally different, through supergauge transformations. I quote Wess' recollection 
{\sl From symmetry to supersymmetry} published in \cite{KS}, p. 67:
\begin{quote}
Another path to supersymmetry came from two-dimensional dual models.
Neveu\index{Neveu} and Schwarz had constructed models which had
spinorial currents related to supergauge tran\-sformations that
transform scalar fields into spinor fields. The algebra of the
transformation, however, only closed on mass shell. The
spinorial currents were called supercurrents and that is where the
name ``supersymmetry'' comes from.

In 1974 Bruno Zumino\index{Zumino} and I published a paper \cite{WZ} where
we established supersymmetry in four dimensions, constructed
renormalizable Lagrangians and exhibited nonrenormalization
properties at the one-loop level.
\end{quote}

CERN was the right place for a discovery of such caliber. It had a large and vibrant Theoretical Division, the idea was picked up by many 
eager enthusiasts and quickly spread all over the world, especially in Europe. 
Thus,  Wess and Zumino started an avalanch. The next step which put the construction of supersymmetric models on an ``industrial" basis was the invention of superspace and superfieeds by Salam amd Strathdee \cite{SS2}. In addition to four conventional space-time dimensions they added four ``quantum" dimensions parametrized by Grassmann numbers,
\beq
\{t,x,y,z\} \to \{t,x,y,z,\theta_\alpha,\bar\theta_{\dot\alpha}\} 
\eeq
where the Grassmann parameters $\theta_\alpha,\bar\theta_{\dot\alpha}$ anticommute and, therefore, say $\theta_\alpha\,\theta_\alpha\equiv 0$ for $\alpha=1,2$.
Then superfields depend on $x_\mu$ and on $\theta_\alpha,\,\,\bar\theta_{\dot\alpha}$ while their various $\theta$ components are regular fields.
 \begin{figure}[t]
\centerline{\includegraphics[width=6cm]{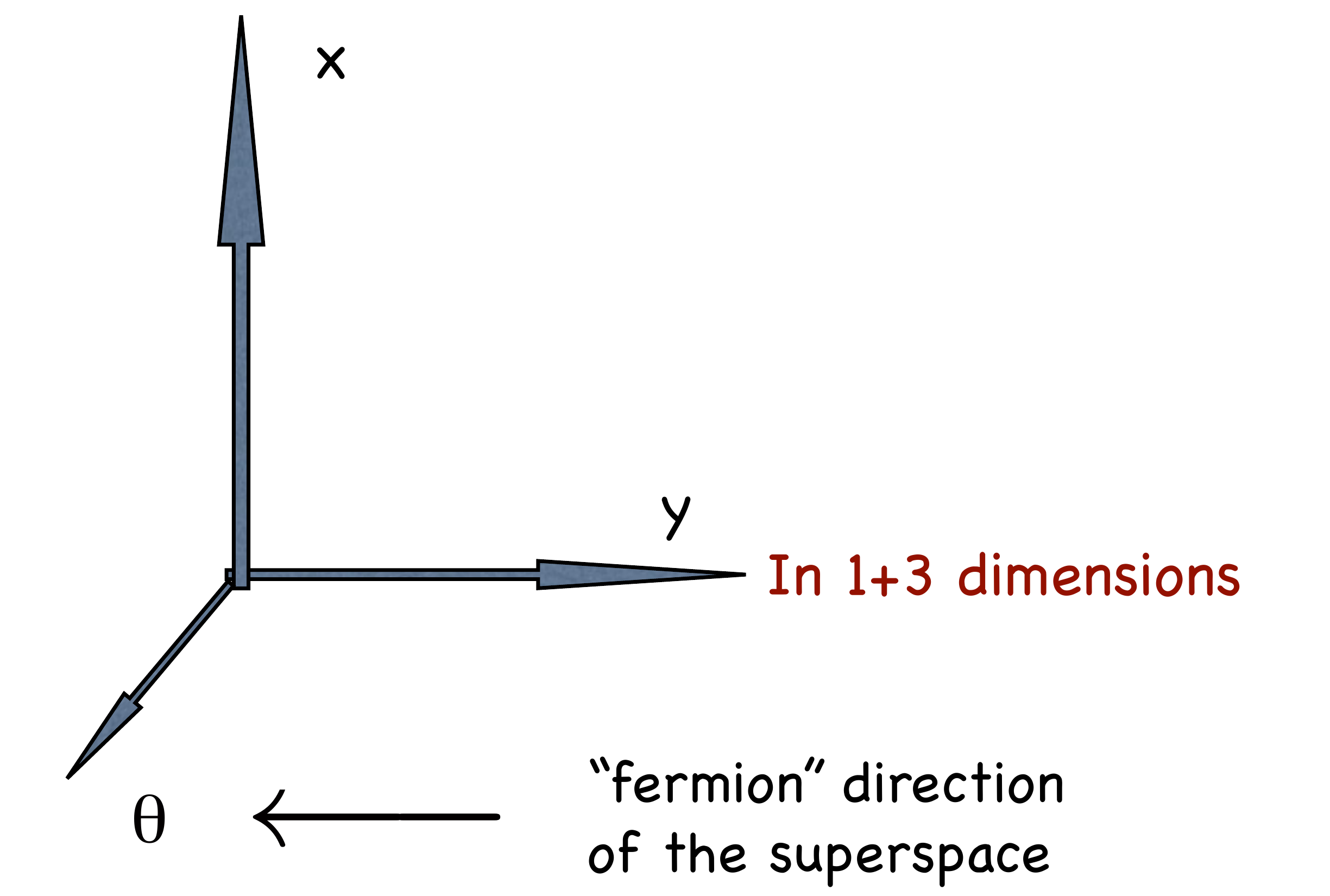}}
\caption{\small 
A simbolic depiction of superspace in 1+3 dimensions.
} 
\label{supers} 
\end{figure}

In the late 1970s, after basic theoretic structural elements were completed,  a number of people turned to supersymmetry-based phenomenology. The trend grew like a snow-ball after the famous 1982 Witten's paper \cite{wit}. If you look at Inspire-hep or Google scholar you will find tens of thousands of papers on supersymmetry-based phenomenology written in the next 20 or 25 years.

Why so many theorists embraced supersymmetry-based phenomenology? I see three major reasons and an argument,  related to the notion of naturalness which we will discuss shortly. 
First, supersymmetry is esthetically attractive. Second, 
 supersymmetry promised a natural explanation of the mass hierarchies.
 
  If we set the quark and lepton masses around their experimental values, loop corrections will tend to drag them to higher values. For fermions this is not an absolutely  dramatic effect because chirality protects fermion masses making the loop contributions only logarithmically dependent of the UV cut-off, usually assumed to be the Planck mass,
\beq
M_{\rm Pl} =\frac{\hbar c}{G_N}\sim 10^{19}\,\,{\rm GeV}.
\eeq
Therefore, the fermion mass hierarchy can be naturally organized. The boson masses are quadratically divergent in the UV and, as a result, are dragged all the way up to $m^2_{\rm bosons}\sim M^2_{\rm Pl}$. In the bosonic sector an incredibly precise fine-tuning must be organized to reproduce the observed hierarchy. 

Supersymmetry helps. Indeed, if SUSY was unbroken, the boson-fermion degeneracy will keep their masses together. Of course, in our world supersymmetry is broken. If superpartners are relatively light, say in the ballpark of 
\beq 
M_{ {\cancel{\rm SUSY}}}\sim {\rm  a \,\, few\,\,  hundred \,\,GeV},
\label{7}
\eeq
the Planck mass as the UV cut off for the boson masses will be replaced by $M_{ {\cancel{\rm SUSY}}}$. The most problematic particle in this respect is the Higgs boson because its coupling to the $t$ quark is of the order of one. Until the end of 2020s it was believed that light superpartners will come to the rescue of  the mass hierarchy. 
Well,... this did not happen.

We are aware of another quantity which currently requires even a more extreme fine-tuning -- the vacuum energy density. A natural estimate for the latter would be
\beq
\left({\mathcal E}_{\rm vac} \right)^{\textstyle\frac{1}{ 4}}\sim M_{\rm Pl}\,.
\eeq
As we already know, unbroken supersymmetry implies ${\mathcal E}_{\rm vac} =0$. With broken supersymmetry one can expect  $$\left({\mathcal E}_{\rm vac} \right)^{\textstyle\frac{1}{ 4}}\sim M_{ {\cancel{\rm SUSY}}}\,.$$ Since according to experimental evidence $$\left({\mathcal E}_{\rm vac} \right)^{\textstyle\frac{1}{ 4}}\sim 10^{-3} {\rm eV}$$ even if $M_{ {\cancel{\rm SUSY}}}\sim$100 GeV the discrepancy is $\sim 10^{14}$, fourteen orders of magnitude. The situation with ${\mathcal E}_{\rm vac} $ currently seems  hopeless. It is highly likely that we (I mean theoretical physics 
as a whole) are not ripe enough to address the issue for the time being.

Third, in SUSY models there was a natural candidate from dark matter -- the lightest superpartner (LSP). In SUSY models with the $R$ parity conservation it is stable. 
Its other features and characteristics could be made compatible with what is needed for dark matter candidates.

Finally, an additional argument that I have mentioned above is the fact that the gauge coupling unification is much better in the presence of supersymmetry 
than without it (see e.g. Fig. 3 in Ref. \cite{LP} and references therein).

\begin{figure}[t]
\centerline{\includegraphics[width=9cm]{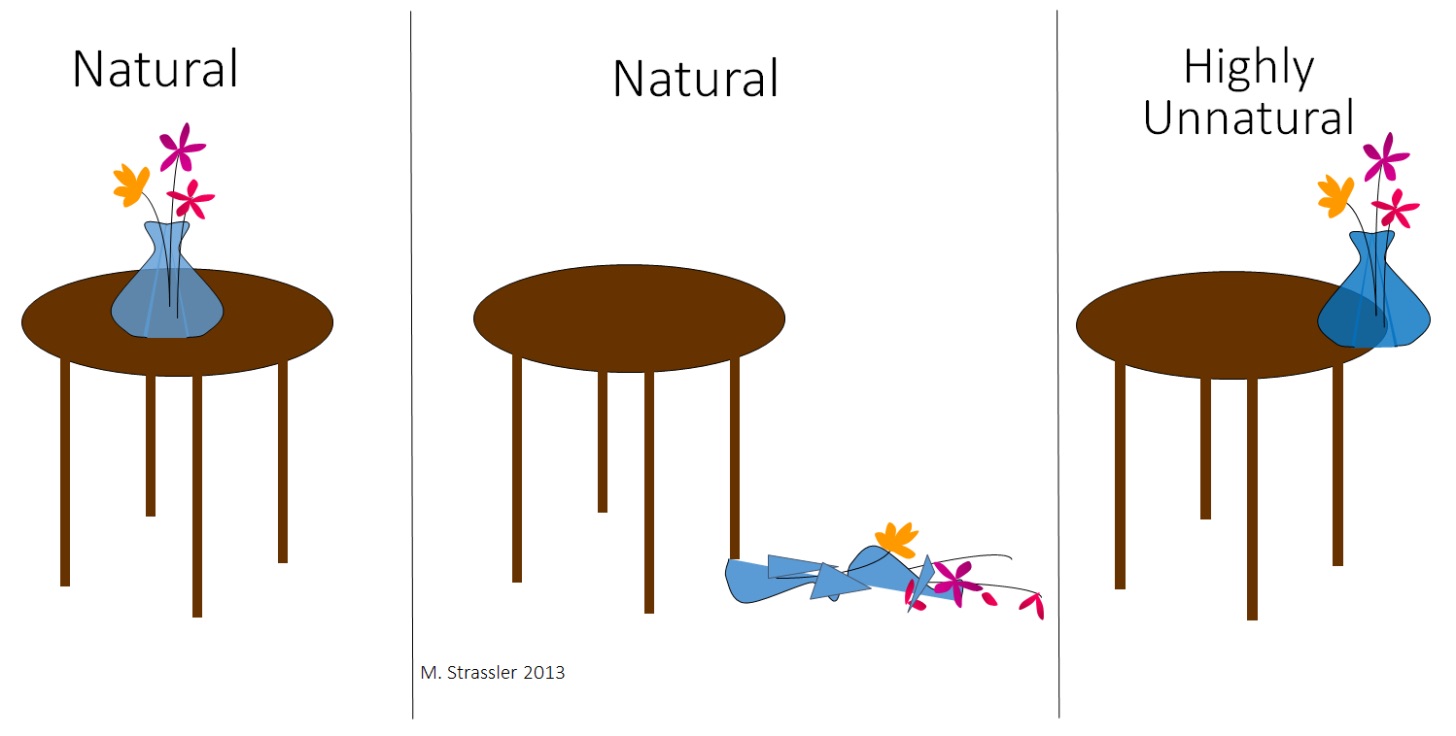}}
\caption{\small 
A symbolic depiction of natural vs. unnatural theories. Courtesy of M. Strassler.
} 
\label{natur} 
\end{figure}

The notion of naturalness became central in high energy theory after the famous 't Hooft's lecture \cite{tho}.
It can be formulated in a number of ways. For instance, a quantity in nature can be small only if the underlying theory protects it by a relevant symmetry. Or, alternatively,  dimension{\em less} ratios between dimensional physical parameters (e.g. mass scales) appearing in a natural  theory should be of order 1,  fine-tuning ``by hand,"
without a visible reason,  is unnatural (see Fig. \ref{natur}). 

Supersymmetry attracted so much attention in particular, because it promised to explain naturalness of the  Higgs mass. Alas... all our expectations failed.  So far LHC produced no evidence of superpartners in the expected mass range, and no evidence of dark matter candidates, not even hints. Thus the huge mass hierarchy we observe in nature remains incomprehensible, let alone the incredible smallness of the vacuum energy density.

What should be done under the circumstances?

Of course, one can still hope that  new breakthrough discoveries will be made at a later stage of the LHC operation.  But I think currently it is reasonable to forget
about naturalness and learn how to live without it. In due time we will receive a hint from heavens which will reveal a proper direction of thought in the issue of
${\mathcal E}_{\rm vac}$ but this time has net yet come.

As a substitute for the naturalness paradigm, a Multiverse conjecture was put forward (also known as {\em landscape}\footnote{The term landscape comes from the notion of a fitness landscape in evolutionary biology. It was first applied to cosmology by Lee Smolin in his 1997 book \cite{Lsm}, and was first used in the context of string theory by Leonard Susskind \cite{suss}.}). Replacing our Universe, Mutiverse is supposed is supposed to consist of a huge number of 
universes, perhaps $10^{\sim 500}$ (infinite for all practical purposes)  which are causally disconnected from each other. This philosophic concept emerged in string theory
in which each particular compactification produces a different spacetime, interpreted as a ``universe." These parallel universes are expected to have different physical parameters, but the same basic laws. The hope is that among the multitude of worlds accidentally there are few with the set of parameters allowing human-like life to develop. 
Steven Weiberg, one of the first and greatest proponents of the anthropic principle \cite{SW}, was very enthusiastic about the Multivrse/landscape idea justifying it by the following remark:
\begin{quote}
 ...[T]he hope of finding a rational explanation for the precise values of quark masses and other constants of the standard model that we observe in our Big Bang is doomed, for their values would be an accident of the particular part of the multiverse in which we live \cite{SW1}.
\end{quote}
Landscape may or may not exist -- at the moment it is a philosophical (or religious, if you want) rather than physical issue. To my mind, a physical theory needs to be falsifiable in order to be scientific. As far as I understand, there is no experimental way to prove or disprove the Multiverse concept. 

Let me summarize this part of the talk. Naturalness was a belief.  
It is not something imposed by existing data/observations. So was the belief that string theory is the 
ultimate theory of everything. Both believes lead us to a dead end. Attempts to skip experiment are punished by nature which
no longer reveals its secrets. 

Does it mean that SUSY-based phenomenology  or string theory should not 
have been throughly studied?

Absolutely not! Remember the history of Yang-Mills theory? It was discovered in 1954 with the aim of describing the family of vector $J^P=1^-$ mesons (e.g. $\rho$ mesons). It turned out useless for this purpose. However, 20 years later it became a basis for quantum chromodynamics (QCD). What a triumph!  If ambitions are scaled down, from ultimate theory of everything to less global goals, one can say that both supersymmetry and string theory have already brought a lot of fresh ideas in our understanding of strong coupling dynamics in field theory. And this is a great achievement. Their mathematical structure is rich and promising.

The field theory was born in the 1920s-early 1930s. The first 50 years of QFT was devoted mainly to  weak coupling analysis, development of perturbation theory and studies symmetry-based  consequences. These were the years of the triumph of QED, absolute success of the Standard Model  and asymptotic freedom of QCD. Since the 1970s the theoretical focus shifted and QFT gradually became the realm of strong coupling both in HEP,
 Condensed Matter, and even gravity at distances $L\sim \left(M_{\rm Pl}\right)^{-1}$
  (e.g. quantum regime in black holes). Till 1980s QFT at strong coupling advanced much more modestly than in weak coupling. This situation started changing with the advent of supersymmetry. It turned out that SUSY is such a powerful symmetry that it paved the way to exact solutions of some theories in 3D and 4D -- this was never anticipated. The main feature distinguishing SUSY theories from non-SUSY is the fact that, say, in super-Yang-Mills the coupling constants can be complexified which results in holomorphic behavior in certain protected sectors of the theory. Before SUSY, exact results were obtained only in a handful 2D models, such as the Schwinger model, sine-Gordon, etc. To move on to  strong coupling 4D  theories we badly needed a novel powerful tool, and supersymmetry provided us with it. Why did we want to delve in strong coupling? In HEP the main reason for that was a remarkable property of quark confinement.
      
   In the vast majority of physical phenomena interactions
between two interacting bodies fall off with distance at large distances. That's why our world, us included, can exist! On the other hand, interaction between quarks grows linearly with distance;
therefore they are permanently confined forming protons and neutrons which, in turn, give rise to nuclei and simultaneously to all of nuclear physics and quantum chemistry.
Without quark confinement life as we know it would not be possible. We see that understanding confinement is absolutely crucial.
 \begin{figure}[h]
\centerline{\includegraphics[width=4cm]{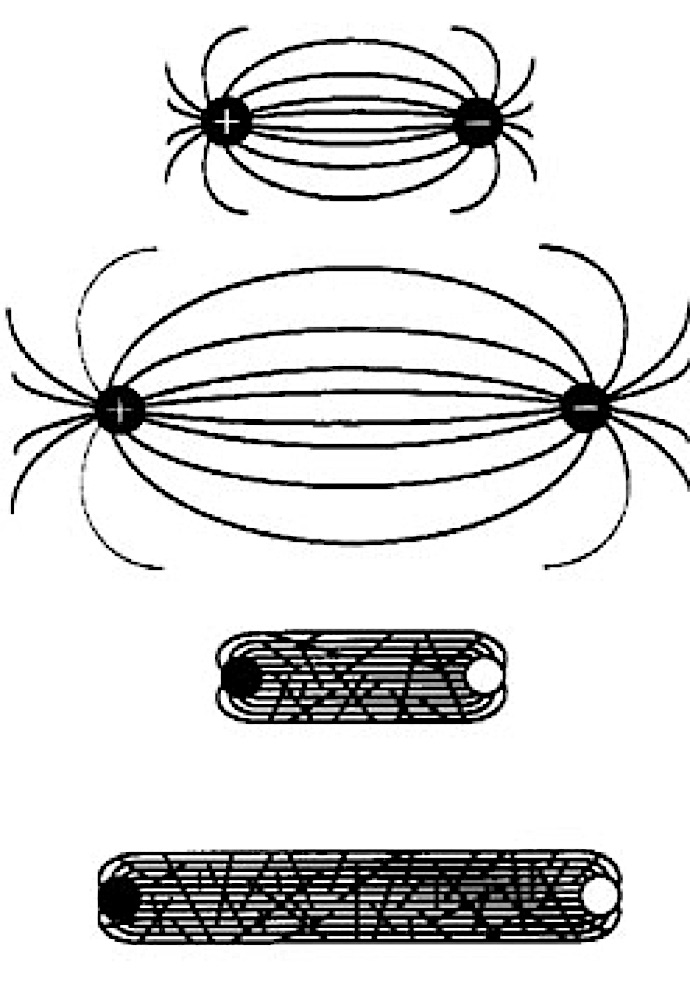}}
\caption{\small 
Coulomb interaction between electric charges vs. confining interaction between quarks.
} 
\label{conf} 
\end{figure}

Figure \ref{conf} illustrates a qualitative difference of  two energy patterns -- the Coulomb interaction between electric charges (interaction energy falls off as $1/R$ where $R$ is the distance between them) vs. confining interaction which grows linearly with distance, i.e. $V=\sigma R$ where $\sigma$ is the flux tube tension. The picture at the bottom of Fig. \ref{conf}  displays a flux tube with a fixed tension which is formed between the quark and antiquark.
 Quarks do not exist in isolation. In experiment we observe only 
quark-antiqurak pairs in the color-singlet state (mesons) or quark triplets (baryons), also in the color-singlet state.

Are we aware of other {\em fundamental}  physical phenomena in which interaction energy between two interacting bodies grows with distance at large distances?
Until the discovery of confinement in QCD there was only one example -- magnetic monopoles in type-II superconductor. Of course, there are no magnetic monopoles we
are aware of. However a very long and thin magnet can be a good substitute in the case at hand. If we have a large sample of a type-II superconductor and 
two such magnets, and we insert the North pole of one magnet into the sample while the South pole of the other from the opposite side of the sample, a magnetic flux tube will form between them which will result in the linear growth of the potential. Such flux tubes were theoretically predicted by Abrikosov \cite{abr} 65 years ago. He also predicted the existence of type-II superconductors.
 \begin{figure}[t]
\centerline{\includegraphics[width=12cm]{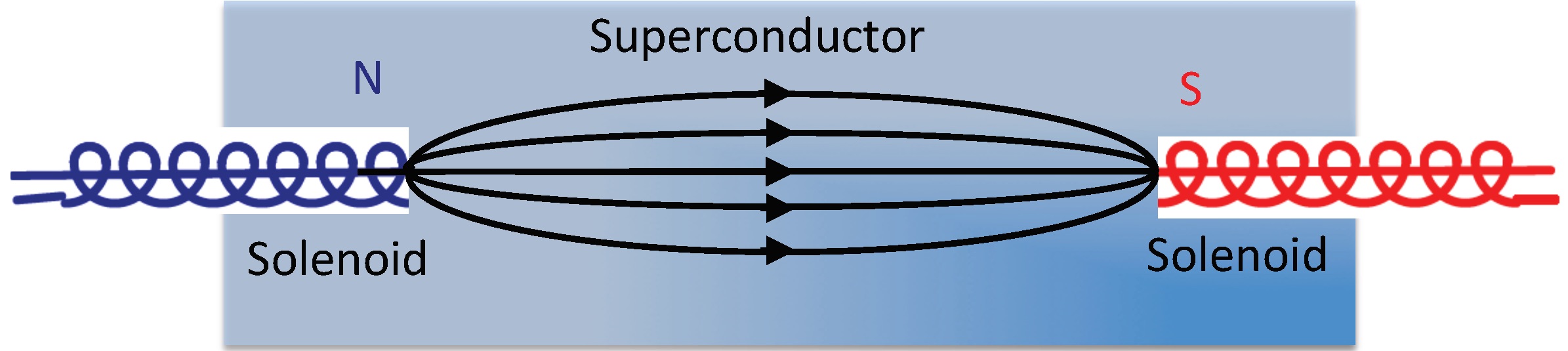}}
\caption{\small 
Abrikosov vortex inside type-II superconductor. The magnetic flux is squeezed into a tube.
} 
\label{conf} 
\end{figure}

A genuine understanding of any phenomenon requires  a developed theory -- not just a numerical calculation -- we need an analytic theory. A theory of quark confinement was badly needed. 

 \begin{figure}[h]
\centerline{\includegraphics[width=5cm]{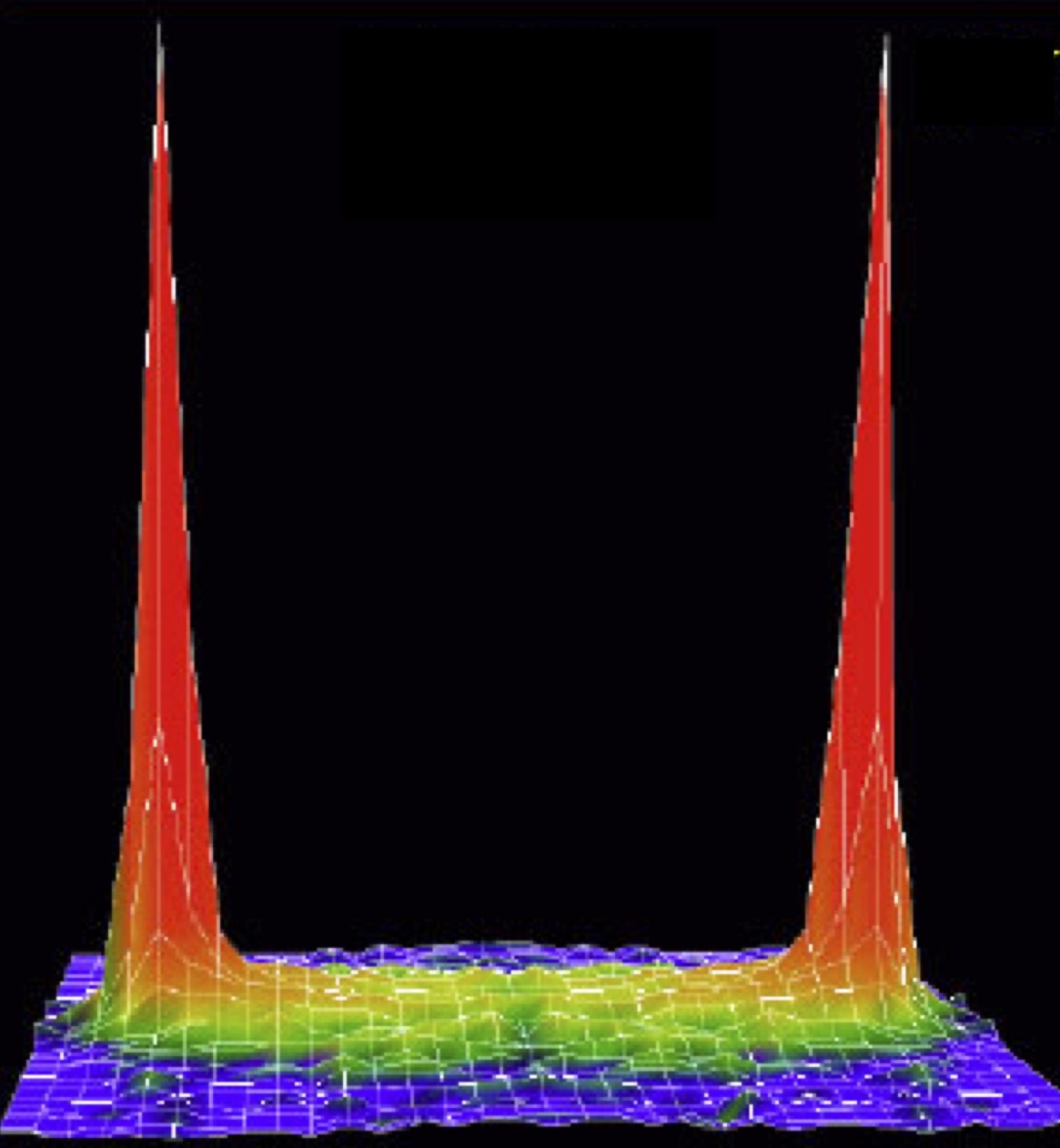}}
\caption{\small 
Lattice simulation of $Q\bar Q$ interaction in SU(2) QCD. Shown is the action density. The interquark distance is 2 fm. We see a tube stretching between $Q$ and $\bar Q$, but have no clue as to the  underlying mechanism for its formation. 
} 
\label{conf} 
\end{figure}

\begin{figure}[h]
\centerline{\includegraphics[width=9cm]{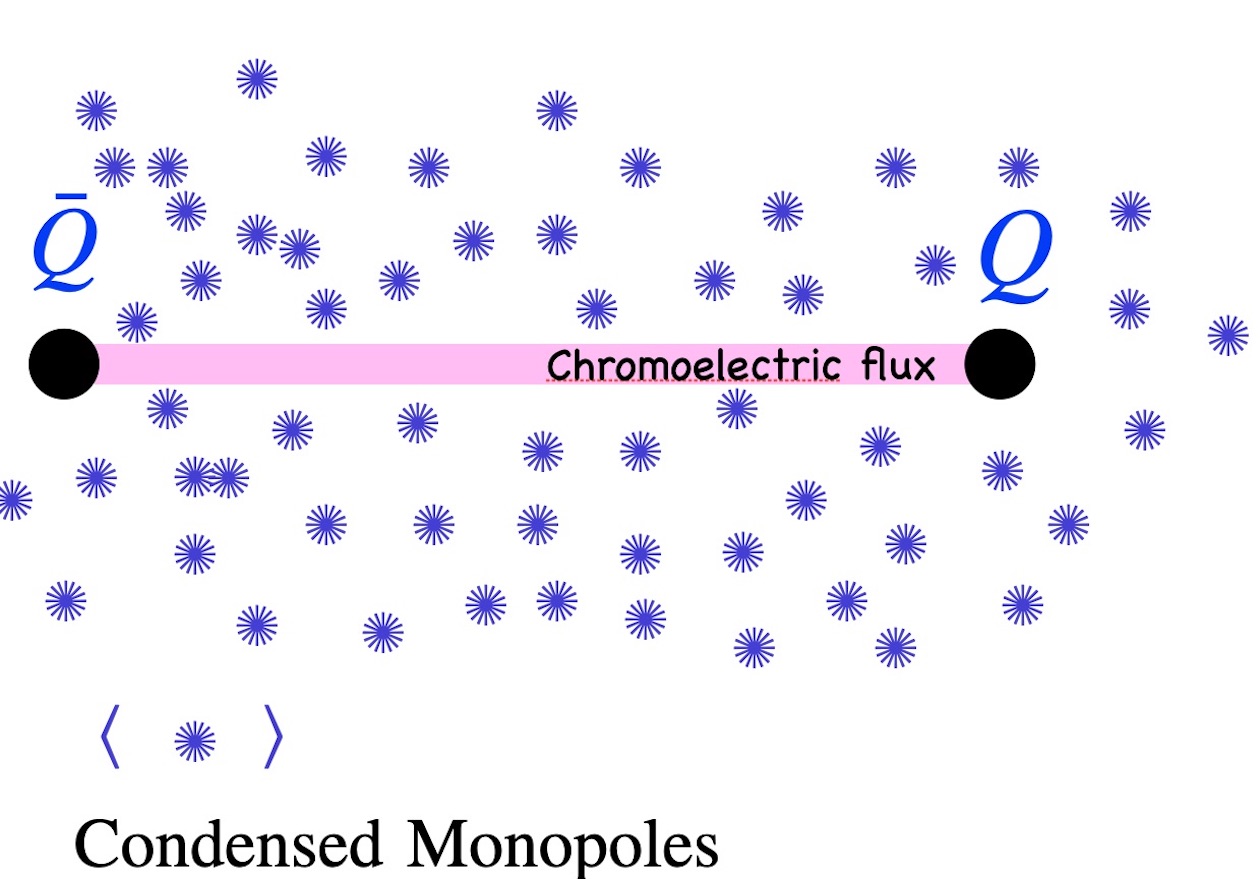}}
\caption{\small 
The dual Meissner effect (see page \pageref{mei}). Condenced QCD ``monopoles" squeeze chromoelectric flux in a tube.
} 
\label{hyp} 
\end{figure}

In 1974-75 an idea was put forward by 't Hooft, Nambu and Mandelstam (independently) \cite{HNM} which stayed a pure conjecture -- or, better to say, a dream -- for 20 years. It goes under the name of the dual Meissner effect. Repulsion of the magnetic field from superconductors was discovered by W. Meissner about a century ago. The Abrikosov vortices are the consequence of this effect. The magnetic flux instead of spreading over the superconducting sample is squeezed into a thin tube which, being attached to monopoles, confines them.
In QCD we deal with quark confinement. Quarks carry chromoelectric charges; therefore, the tube attached to them must carry the chromoelectric flux. That's why we need to understand whether chromoelectric vortices can emerge in QCD at strong coupling. The 't Hooft-Nambu-Mandelstam hypothesis is illustrated in Fig. \ref{hyp}.
While Nambu and Mandelstam's publications
are easily accessible, it is hard to find the EPS Conference Proceedings
in which 't Hooft presented his vision. Therefore, the corresponding passage from his talk
is worth quoting: ``...[monopoles] turn to develop a non-zero vacuum expectation value.
Since they carry color-magnetic charges, the vacuum will behave like a superconductor
for color-magnetic charges. What does that mean?
Remember that in ordinary electric superconductors, magnetic charges are confined by 
magnetic vortex lines ... We now have the opposite: it is the color charges that 
are confined by electric flux tubes."

20 years that elapsed after  't Hooft, Nambu and Mandelstam publications \cite{HNM} people tried to quantify this idea, to no avail! The main difficulty was to formalize the very notion of monopole in the confining phase of Yang-Mills, let alone condensed monopoles. (In the Higgs phase it was easy -- the famous t' Hooft-Polyakov construction \cite{thp} does the job.) 

We approach the crescendo in this symphony. The community had to wait for 20 year before nonperturbative supersymmetry was developed and was used in 1994 by Seiberg and Witten \cite{SeiW} to find an exact analytic solution of the confinement phenomenon in a supersymmetric Yang-Mills theory. Strictly speaking,  they dealt  not with QCD but a SUSY theory which can be viewed as a cousin of QCD. Probably, not the first but rather the second cousin. The Seiberg-Witten theory has an extended SUSY (the so-called 
${\mathcal N}=2$)  -- eight supercharges rather than 4 in the minimally supersymmetric Yang-Mills. The field content is as follows. Let us compare  with Eq. (\ref{2}) presenting the minimal SUSY with four supercharges. It is almost QCD! Well...  ${\mathcal N}=1$ SUSY is not strong enough, and only a few islands in this theory were explored, such as the number of vacua, the exact value of the gluino condensate, etc. It  is quite  upsetting that we cannot go further. 

Supersymmetric gluodynamics  has gluon and gluino as in Eq. (\ref{2}). To pass to  ${\mathcal N}=2$ we must add the second gluino with its superpartner, a scalar complex filed $\phi^a$ in the adjoint representation of the gauge group. The field $\phi^a$ has a flat direction.
Any point in the complex plane Tr$\phi^2$ can be a vacuum. Since Tr$\phi^2$ is complex, certain protected functions are analytic functions of this complex parameter. This is where the power of holonomy enters. 
If $|\phi^a|_{\rm vac}\gg \Lambda$ where $\Lambda$ is the dynamical scale of the theory then we are at weak coupling, in the quasiclassical regime. If $|\phi^a|_{\rm vac}\sim \Lambda$ we are at strong coupling in the highly quantum regime. 

Now one can use analytic continuation to connect  the weak and strong coupling regimes. At weak coupling, when $|\phi^a|_{\rm vac}\gg \Lambda$, the SU(2) gauge symmetry is spontaneously broken down to U(1). Out of three gauge bosons of the SU(2) gauge theory two become Higgsed and heavy (like $W^\pm$ bosons).
The U(1) gauge boson remains massless. This is exactly what happens in the Georgi-Glashow model to which the 't Hooft-Polyakov monopoles are inherent. In the quasiclassical regime they are very heavy; their mass is $\sim M_{W^\pm}/\alpha$.

This was the first step. We have a theory which is rather similar to QCD but has well-defined heavy monopoles. Very heavy objects cannot condense in the ground state. They have to be massless in order to condense. Seiberg and Witten applied holonomy to determine a pre-potential (i.e. the kinetic term) and as a result were  able to identify two special points in the Tr$\phi^2$ plane (two strongly coupled vacua) in which the monopoles become exactly massless. These points are referred as Seiberg-Witten points. Most importantly, they can be continuously analytically connected to heavy 't Hooft-Polyakov monopoles, so even at strong coupling they continue to be well defined. Massless fields are ``ready" to condense. A small deformation of the ${\mathcal N}=2$ theory by an 
${\mathcal N}=1$ operator makes them to condense.

The Seiberg-Witten publication \cite{SeiW} was a breakthrough achievement which reinvigorated theoretical efforts to understand confinement in Yang-Mills theory.
Since I do not have much time left, I will say just a couple of explanatory remarks on  their strategy and  why SUSY played such an instrumental role. The crucial element of SUSY 
is complexification of  certain parameters, e.g. the gauge couplings, with ensuing holomorphy of certain correlation functions  in the so-called protected sectors of the theory,
a phenomenon which has no analogs in non-SUSY theories. To find a holomorphic function is much easier than to find just a function. It is sufficient  to find the position and nature of all singularities of holomorphic functions to determine the function {\em per se}. The brightest idea of Seiberg and Witten was to use the extended ${\mathcal N}=2$ holomorpy to fully determine the  {\em pre-potential} in the low-energy limit in which SU(2) Yang-Mills reduces to ${\mathcal N}=2$ SUSY QED. 
To see what happens at strong coupling Seiberg and Witten 
 dualized the original SUSY QED. Everythinng magnetic is converted in electric and {\em vice versa}, $\vec E \leftrightarrow\vec B$. The ``quarks" of the dualized QED are the ``monopoles" of the original one. Thus, condensation of quarks in the dualized theory is equivalent to condensation of monopoles in the original theory with the subsequent formation of ``chromoelectric" flux tubes.
 
 The strong coupling regime in the original SUSY QED corresponds to weak coupling regime after dualization of the theory in which everything becomes calculable. In this way it was  analytically demonstrated, for the first time ever, that the dual Meissner effect \label{mei} is indeed implemented in the Seiberg-Witten theory through formation of chromoelectric flux tubes. A triumphant confirmation of the 't Hooft-Nambu-Mandelstam hypothesis was obtained.
 The golden dream came true!
 Well... not quite. Our genuine golden dream is to obtain this phenomena in ${\mathcal N}=0$, i.e. nonsupersymmetric Yang-Mills.

\vspace{2mm}

\begin{figure}[h]
\centerline{\includegraphics[width=6cm]{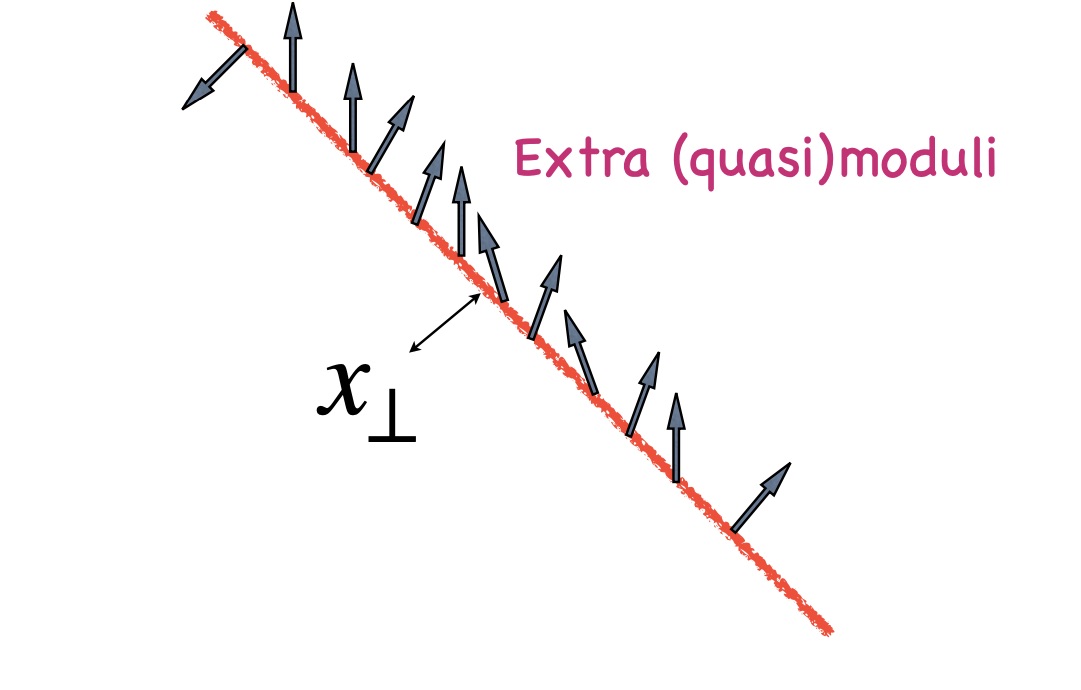}}
\caption{\small 
Non-Abelian string with a global SU($2)$ symmetry.
} 
\label{nast} 
\end{figure}

The confining string obtained by Seiberg and Witten is of the Abrikosov-Nielsen-Olesen  type \cite{ANO}.  In fact, the dual SUSY QED in the bosonic sector coincides with  the model considered by Abrikosov, Nielsen and Olesen. To advance closer to QCD the so-called non-Abelian string was invented. Its characteristic feature is 
that {\em all} non-Abelian degrees of freedom equally participate in dynamics at the scale of string formation.  It can be obtained from the Seiberg-Witten model by adding $N_f$ matter fields in the fundamental representation 
with the number of flavors coinciding the number of colors, $$N_f=N\,,$$ and adding an extra photon field endowed with the Fayet-Iliopoulos term \cite{Yung,Yung2}. In this model color-flavor locking takes place, namely both, the gauge SU($N)$ symmetry and the global SU($N_f=N)$ are spontaneously broken but their diagonal SU($N)$ subgroup survives as a global symmetry. All gauge bosons are Higgsed and acquire the same common mass. Symbolically this string is depicted in Fig. \ref{nast}
For simplicity I will assume that $N=2$.
Then the unit vectors $\vec S(t,z) $ depicted in Fig. \ref{nast} are attached  to every point of the string. They appear as moduli (collective coordinates) in addition to the translational moduli $x_\perp$ which fully characterize the Abrikosov string.
The extra moduli $\vec S$ are referred to as {\em rotational} since they express the freedom of rotations of the string solution in the SU(2) global group.\footnote{In the case at hand we have two rotational moduli. Indeed three-dimensional unit vector is fully parametrized by two angles.} The unit vectors acquire their own two-dimensional dynamics. Their interaction is described by the Heisenberg model 
\beq
{\mathcal L}_{\rm Heis} \sim \partial_a\vec{S} \, \partial^a
\vec{S}+{\rm fermions}, 
\label{heis}
\eeq
where $a=0,1$. If supersymmetry of the Seiberg-Witten model is ${\mathcal N}=2$  in four dimensions, supersymmetry 
of the Heisenberg model  on the string world sheet is two-dimensional ${\mathcal N}=2$ (four supercharges). The flux tube shown by a red line in Fig. \ref{nast} is chromomagnetic. Unfortunately, its dualization is not  worked out in full.

The above construction exhibits another remarkable 
a phenomenon that owes its existence to supersymmetry and goes under the name of  2D-4D correspondence \cite{Yung2}. What does it mean? 

\begin{figure}[h]
\centerline{\includegraphics[width=6cm]{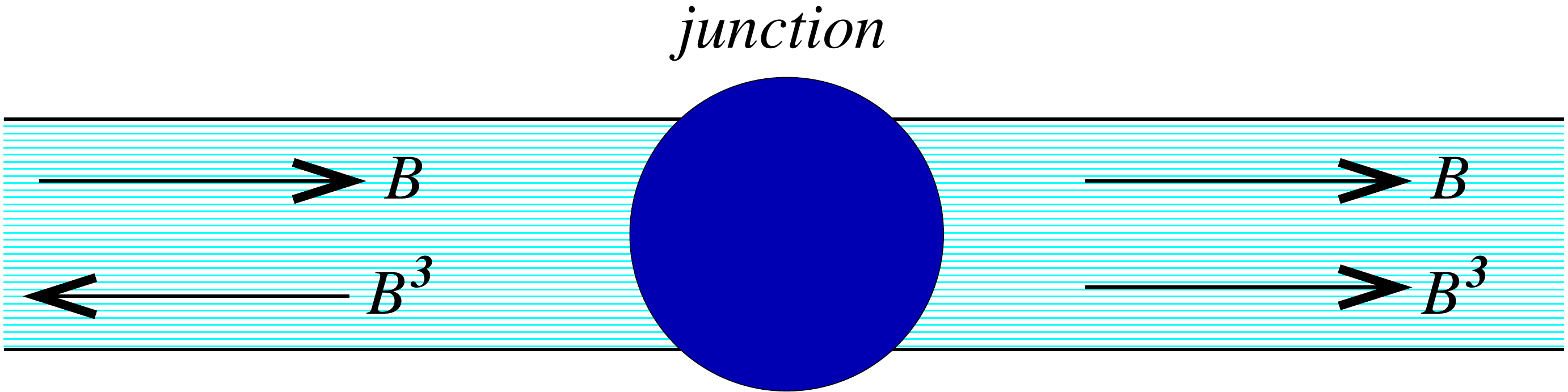}}
\caption{\small 
The junction of two distinct degenerate magnetic flux tubes in SU(2) theory.
} 
\label{dvez} 
\end{figure}

There are two distinct types of the flux tubes in SU(2) Yang-Mills. They are different but their tensions are exactly degenerate, and therefore being exactly degenerate they support a stable junction between them, a transitional domain,  which is called  kink. On Fig. \ref{dvez} on the left is the first type, on the right is the second type. The mass of this kink can be determined by virtue of SUSY. Here comes the 2D-4D correspondence. On the string world sheet we have the model (\ref{heis}) which is strongly coupled in the infrared. However, it is exactly solvable as far as the string tension and kink mass are concerned. 

Now, let us look at the above two-dimensional construction from the 4D perspective.  How can we interpret it?
Again, using the power of holonomy  one can show that it is nothing other than  the analytic continuation of the 't Hooft-Polyakov monopole with two magnetic strings attached to it, a confined monopole. Of coursed, the 't Hooft-Polyakov monopole does not look similar to the kink  but be sure, they are related by analytic continuation. The mass of the 2D kink is given exactly by the same expression as the mass of the 4D monopole,
\beq
M_{\rm conf\,\, mon} \leftrightarrow M_{\rm kink}\,,
\eeq
if expressed in appropriate parameters. This correspondence would be impossible to obtain  without holomorphy in SUSY theories.

\vspace{2mm}

\begin{figure}[h]
\centerline{\includegraphics[width=4.5cm]{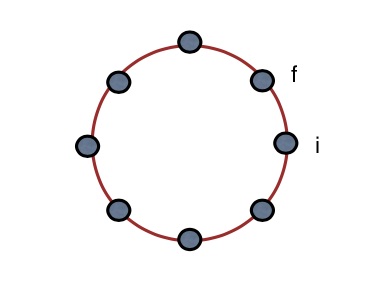}}
\caption{\small 
$N$ vacua in SU($N$) SUSY gluodynamics are shown in the complex plane of the parameter $\langle
\lambda^{a}_{\alpha}\lambda^{a\,,\alpha}
\rangle$. Two neighboring vacua between which the domain wall interpolated are marked by i,\,f.
} 
\label{vacu} 
\end{figure}

Let us return for a short while to minimal ${\mathcal N}=1$ supersymmetry in SUSY gluodynamics (\ref{2}). If the gauge symmetry is SU($N$) this theory has $N$ vacua, 
labeled by the vacuum expectation value of the gluino condensate,
 \beq
2  {\rm Tr} \,\left\langle \lambda^\alpha \lambda_\alpha\right\rangle =
\langle
\lambda^{a}_{\alpha}\lambda^{a\,,\alpha}
\rangle = -6 N\Lambda^3 \exp \left({\frac{2\pi i k}{N}}\right)\,,
\,\,\, k = 0,1,..., N-1\,,
\label{wall15}
 \eeq
 see Fig. \ref{vacu}.
 A special type of domain walls were found, interpolating between the degenerate vacua, for instance, the neighboring vacua  i,\,f \cite{DS}. In fact, 
 they are protected (more exactly BPS protected) and therefore, their tension $T$ is exactly calculable,
 \begin{equation}
T_{\rm wall}  = \frac{N}{8\pi^2} \left| \langle {\rm Tr} \lambda^2\rangle_{\rm vac \,\,f}
-\langle {\rm Tr} \lambda^2\rangle_{\rm vac \,\,i}
\right|.
\label{eqte}
\end{equation}
 As was shown by Witten shortly, 
these domain walls are field-theoretic analogs of $D$ branes \cite{Witten}. The key element is the linear $N$ dependence in (\ref{eqte}) which corresponds to $1/g_s$ in string theory.  One can say they are $D$ branes in flesh and blood. That's another example of the power of SUSY. Is it a full solution of ${\mathcal N}=1$? No, it is not.
Does it add something new to our understanding of the strong coupling dynamics? Certainly, it does. 

\vspace{2mm}

\begin{figure}[h]
\centerline{\includegraphics[width=6cm]{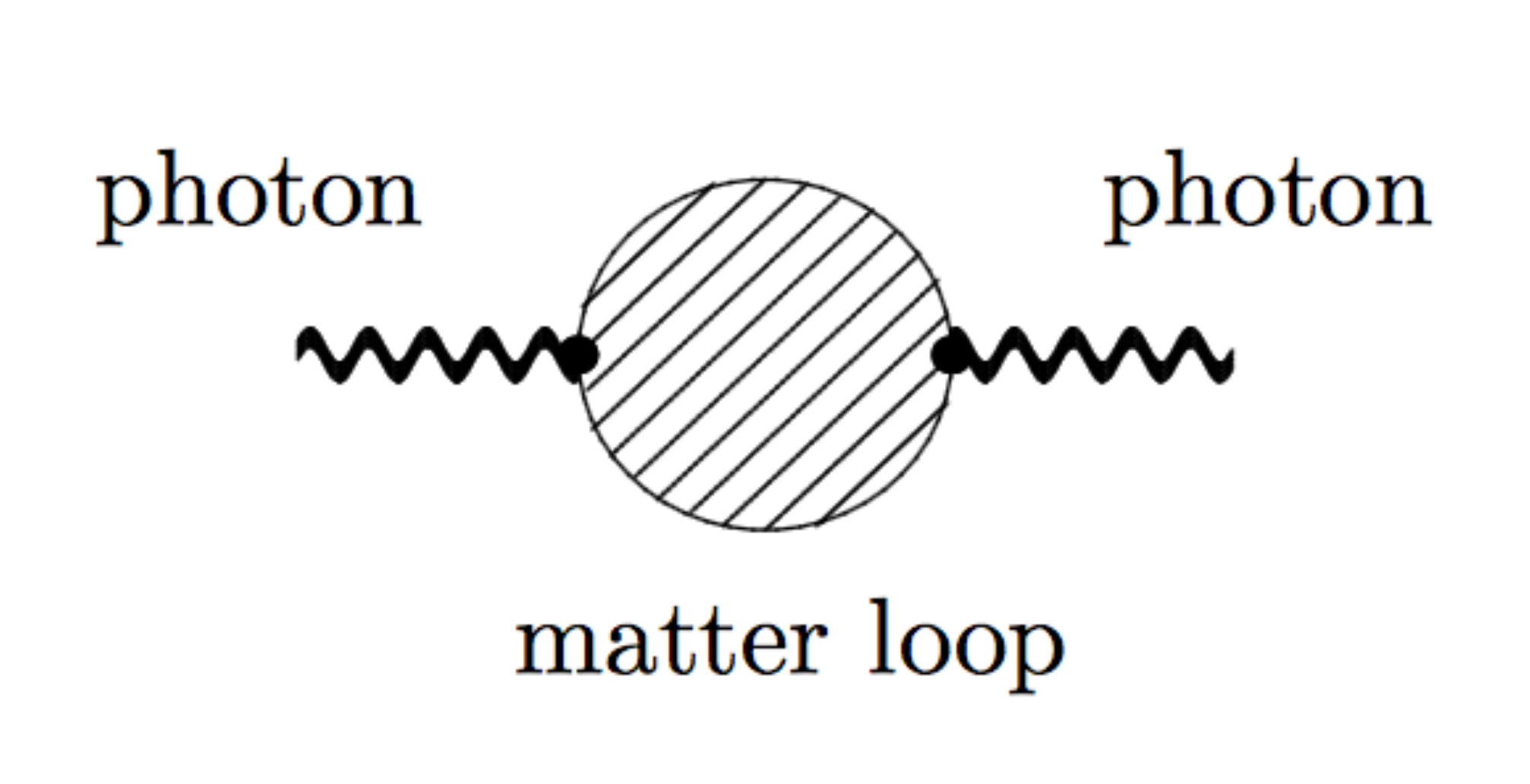}}
\caption{\small 
Two-piont function of virtual photons with the total momentum $q$. If $q^2 $ is positive, its imaginary part is related to $\sigma(e^+e^-\to{\rm quarks, \,\,squarks, \,\, gluons, \,\,and \,\, gluinos})$. Our calculation refers to the Euclidean domain of positive $Q^2\equiv -q^2$.
} 
\label{hcs} 
\end{figure}

In the remaining couple of minutes I'd like to say a few words about a relatively new result which was not reported at previous conferences -- a supersymmetric analog of $e^+e^-$ annihilation in hadrons. Since the focus is on strong interactions (SUSY QCD) we can truncate $e^+e^-$ and consider the two-point function of (virtual) photons with total Euclidean momentum $Q^2$. Its imaginary part gives the hadronic cross-section, see Fig. \ref{hcs}. We will combine SUSY QCD with SUSY QED. The number of {\em massless} flavors is $N_f$. The exact formula for the Adler function $D(Q^2)$ in this theory was obtained in \cite{MSKS},
\beq
D(Q^2) = \frac 32 N_c \sum_f q_f^2 \Big[1- \gamma (\alpha_s (Q^2))\Big]
\label{tsix}
\eeq
where the sum runs over all flavors, $q_f$ is the electric charge of a given flavor, $\gamma$ is the anomalous dimension of the matter
fields. It is the same for all matter fields assuming that they all belong to the fundamnental representation of color. Moreover, $\alpha_s$ is the strong coupling constant.
Note that all complexities of the strong interactions reside in a single function --  the anomalous dimension $\gamma (\alpha_s \left(Q^2\right))$. Although they are not BPS protected and hence are not exactly calculable some general features are known from the NSVZ $\beta$ function \cite{nsvz}. Namely, for each value of $N$ and $N_f$ in a certain interval the $\beta$ function has to zeroes, corresponding to two fixed points: one in the UV, at $\alpha_s=0$ and another in the IR, at 
\beq
\gamma (\alpha_s) \to - \frac{3N-N_f}{N_f}\,.
\label{14}
\eeq
The point $3N=N_f$ is the upper edge of the {\em conformal window}. At this point the UV and IR conformal points coincide.\footnote{It is important that to the leading order
$ \gamma (\alpha_s) =-\left(N-\frac{1}{N}\right) \,\frac{\alpha_s}{2\pi} $. Therefore, if $\gamma =0$ the coupling constant $\alpha_s$ vanishes.} The lower edge of the conformal window can be obtained from the dual ``magnetic" theory (see \cite{seib}) in which
\beq
N\to N_f-N\,,\quad \gamma\to \gamma_D = - \frac{2N_f-3N}{N_f}
\label{15}
\eeq
implying that at the lower edge $\frac 32 N=N_f$. The conformal window stretches 
\beq
\frac 3 2 N \leq N_f\leq 3N\,.
\label{16}
\eeq
In the conformal window the RG flow of $\gamma$ is as follows: from 0 at the UV to (\ref{14}) if $N_f\leq 3N$. For the dual anomalous dimension $\gamma_D$ runs from 0 at the UV to (\ref{15}) if $\frac 3 2 N \leq N$. If we consider both the ``electric" and ``magnetic" theories simultaneously,
\beq
{\rm Max} \left\{\gamma,\,\gamma_D \right\} =\frac 1 2.
\eeq
When $N_f$ is close to $3N$ the electric theory is weak, the magnetic is strong, if $N_f$ is close to $\frac 3 2 N$ the situation is reversed. The change of the regime occurs at  $N_f=2N$. At this point in both theories $\gamma^* =\left(\gamma_D\right)^* = \frac 1 2$ where the asterisk marks the limiting value in the IR. To the right the electric theory has a weaker coupling, to the left magnetic.
\begin{figure}[h]
\centerline{\includegraphics[width=9cm]{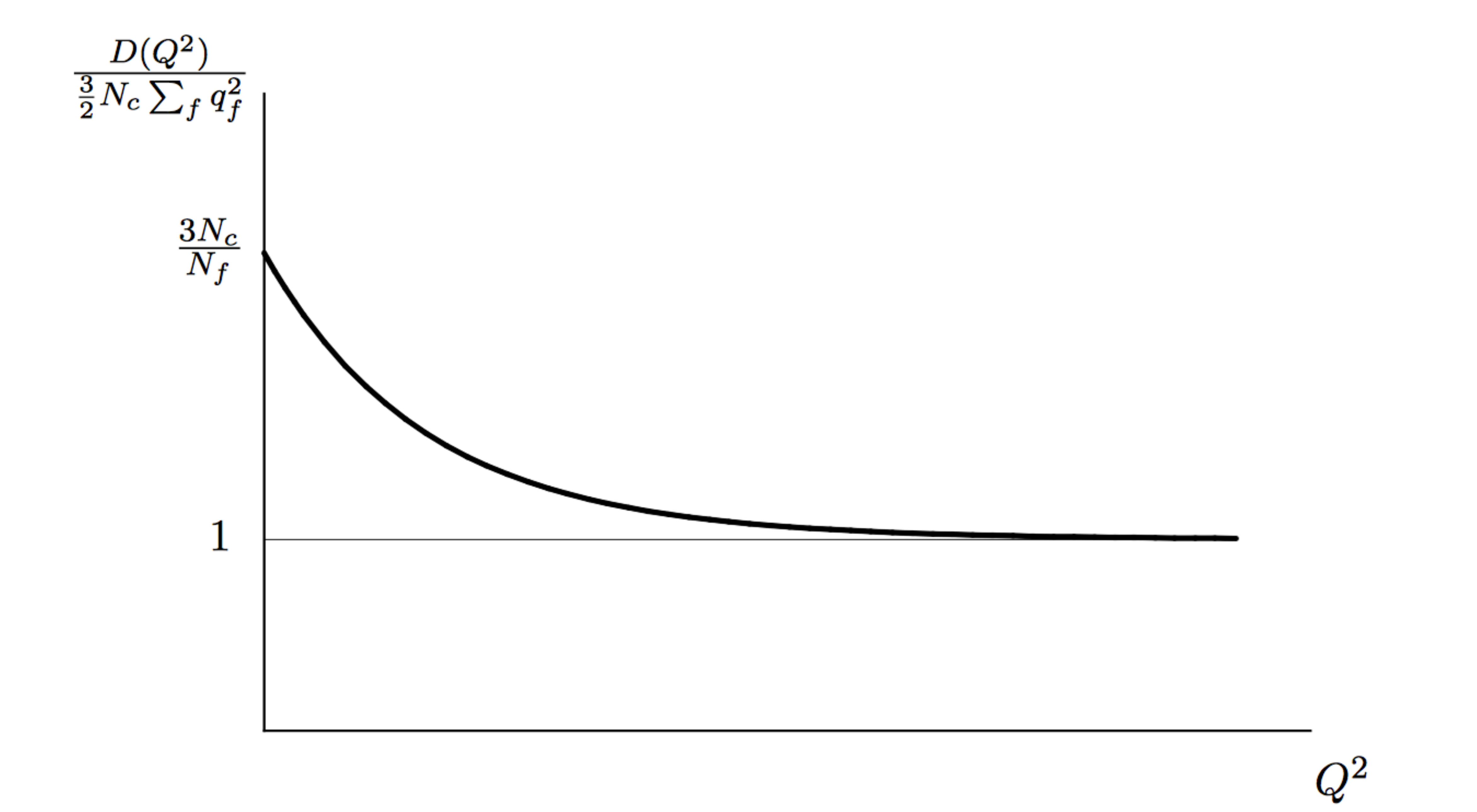}}
\caption{\small 
Two-piont function of virtual photons with the total momentum $q$. If $q^2 $ is positive, the imaginary part of this two-point function is related to 
$\sigma(e^+e^-\to{\rm quarks, \,\,squarks, \,\, gluons, \,\,and \,\, gluinos})$. Our calculation refers to the Euclidean domain of positive $Q^2\equiv -q^2$.
} 
\label{ee} 
\end{figure}
If $N_f > 3N$ the electric theory is IR free (and thus uninteresting). The same happens with the magnetic theory at $N_f< \frac  3 2 N$.
Thus, we will focus on the conformal window (\ref{16}) assuming that $N\to\infty$. Only planar diagrams survive in this limit.\footnote{This set up known as the Banks-Zaks regime \cite{BZ} was discussed recently in a non-supersymmetric context \cite{Shifm}.} Note that in the conformal window the value of $\alpha_s(Q^2)$ is limited from above because of the IR fixed point.

\begin{figure}
\centerline{\includegraphics[width=5.5cm]{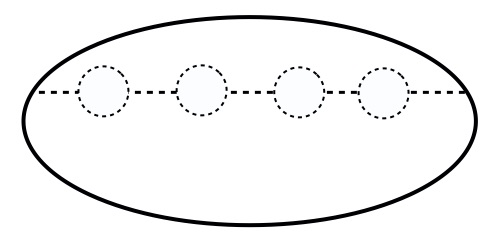}} \caption{Graph showing four loops renormalizing a gluon line (represented by the dotted line). A renormalon is the sum 
over $n$ of such diagrams with $n$ loops.
\vspace{-1mm} } 
\label{peng}
\end{figure}

Let us ask ourself what is the status of the divergence of the $\alpha_s$ series at high orders $k\gg 1$ in the regime under consideration. The number of planar graphs does {\em not} grow factorially \cite{Nussinov}. Instantons are  suppressed as $e^{-N}$ with a positive constant $c$. The only potential source of the factorial growth (as $k!$) of the expansion coefficients are 
renormalons, see Fig. \ref{peng}.

Being treated formally the renormalon graphs, shown in Fig. 3, lead to factorial ($k!$) divergence at high orders. Using 
 Wilson's OPE -- the scale separation  principle -- we eliminate renormalons, or better to say, trade them for a conspiracy between a given renormalon with a certain operator in OPE \cite{Shifm3}. 
The problem is that in SUSY theories  the leading operator, the gluon condensate $\langle G_{\mu\nu}^a  \,G^{\mu\nu\, a}\rangle$ vanishes. This puzzle was addressed 
in the last paper in \cite{Shifm3}, but no final solution was found. The reason is that the standard way of the renormalon graphs isolation is through matter loops while 
in the above paper only pure super-gluodynamics was considered. In \cite{MSKS}
we fully supersymmetrized the standard (non-SUSY) renomalon construction. My conclusion is that in the conformal window we have just discussed there are {\em no} renormalons, i.e. the bubble neckless graphs do not produce factorial divergences in the $\alpha_s$ expansion coefficients because $\alpha_s(Q^2)$ is limited from above and never becomes too strong.
No conspiracy is needed. 

{\em In summary,} SUSY Yang-Mills theories are alive and well. SUSY-based tools at strong coupling provide a unique possibility to address problems
which could not have been solved in any other way. The corresponding results
proliferate.

A large variety of exact results in 4D, 3D and 2D theories, including mass spectrum, exact $\beta$ functions, etc., help us to unravel mysteries of strong coupling in non-supersymmetric theories.

\vspace{3mm}

{\em Acknowledgments}

This work is supported in part by DOE grant DE-SC0011842.

\end{document}